**What is the problem being addressed by the manuscript and why is it important to the Antennas & Propagation community? (limited to 100 words).**

This manuscript addresses the critical problem of strong mutual coupling in a closely spaced stacked microstrip patch antenna (S-MPA) pair, which is a major challenge in compact MIMO and in-band full-duplex (IBFD) systems. Existing decoupling techniques primarily target weak coupling scenarios and are ineffective when inter-element spacing is sub-half-wavelength. The proposed broadband self-decoupling mechanism achieves high isolation without additional decoupling circuits or complex structures, making it highly relevant for next-generation wireless terminals. This work introduces a new paradigm for mutual coupling suppression under co-directional surface currents, with practical impact on the design of compact, high-performance antenna-in-package (AiP) systems.

**What is the novelty of your work over the existing work? (limited to 100 words).**

This work presents the first S-MPA self-decoupling mechanism that achieves broadband high isolation in closely spaced S-MPA elements under the co-directional surface current distributions. Unlike existing decoupling methods that rely on oppositely directed currents and require complex decoupling structures or half-wavelength spacing, the proposed approach uses a simple metallic coupling line between parasitic patches to suppress strong intra- and inter-band coupling without added footprint. It also demonstrates, for the first time, effective self-decoupling between adjacent-band antenna pairs (e.g., N77-N78), offering a structurally simple, fabrication-tolerant, and scalable solution for compact MIMO and IBFD AiP applications.

**Provide up to three references, published or under review, (journal papers, conference papers, technical reports, etc.) done by the authors/coauthors that are closest to the present work. Upload them as supporting documents if they are under review or not available in the public domain. Enter "N.A." if it is not applicable.**

[1]. S.-H. Xing, C. Zhang, M.-Z. Li, X.-Z. Bo, S.-H. Liu and Z.-G. Liu, "Simultaneous enhancing decoupling bandwidth and impedance bandwidth of stacked microstrip patch antennas by achieving the mutual coupling nulls," *IEEE Trans. Antennas Propag.*, (Early Access).
[2]. S.-H. Xing and Z. -G. Liu, "A novel approach to improving isolation of low-profile stacked patch antenna array," *2024 IEEE MTT-S International Microwave Workshop Series on Advanced Materials and Processes for RF and THz Applications (IMWS-AMP)*, 1-3 Nov 2024, Nanjing, China.
[3]. Q. Li, Z.G. Liu, W.B. Lu, "Cascaded Split-Ring Resonator for Decoupling of Broadband Dual-Polarized Base Station Antenna Array," *2021 13th Global Symposium on Millimeter-Waves & Terahertz (GSMM)*, 23-26 May 2021, Nanjing, China.

**Provide at least three references (journal papers, conference papers, technical reports, etc.) done by other authors that are most important to the present work. These references should also be discussed in the submitted manuscript and listed among its references. Please include the citation numbers used in the manuscript for easy reference.**

[1]. M.-N. Wang, Z.-J. Shao, M. Tang, Y. -P. Zhang and J.-F. Mao, "Miniaturization and decoupling of wideband stacked patch antennas based on spoof surface plasmon polaritons," *IEEE Trans. Antennas Propag.*, vol. 72, no 10, pp. 8076-8081, Oct. 2024.
[2]. J.-F. Qian, S. Gao, B.S. Izquierdo, H.Y. Wang, H. Zhou, and H. Xu, "Mutual coupling suppression between two closely placed patch antennas using higher order modes," *IEEE Trans. Antennas Propag.*, vol. 17, no. 6, pp. 4686-4694, Jun. 2023.
[3]. O. Q. Teruel, Z. Sipus, and E. R. Iglesias, "Characterization and reduction of mutual coupling between stacked patches," *IEEE Trans. Antennas Propag.*, vol. 59, no. 3, pp. 1031-1036, Mar. 2011.

# A Self-Decoupling Mechanism for Closely Spaced Stacked Microstrip Patch Antenna Pair with Co-Directional Surface Currents

Shao-Hua Xing, Zhen-Guo Liu, *Senior Member*, *IEEE*, Chao Zhang, Yi-Hao Liu

*Abstract*—This paper presents a simple and cost-effective broadband self-decoupling mechanism to mitigate strong mutual coupling in tightly stacked patch antenna pairs. Unlike conventional decoupling approaches that rely on oppositely directed surface currents between parasitic and driven patches, the proposed method achieves broadband self-decoupling under co-directional surface current distributions by introducing an embedded ultra-narrow metallic coupling line (EUNMCL) between adjacent parasitic patches. This design effectively mitigates boresight gain reduction and total efficiency degradation typically introduced by conventional decoupling techniques, without requiring additional decoupling circuits or complex fabrication processes. In a tightly spaced two-element array, the proposed method enhances isolation by 16.9 dB across the 5G NR N78 band (3.3–3.8 GHz), reaching a maximum improvement of 40.2 dB. It also supports compact adjacent-band MIMO systems, maintaining mutual coupling levels below -20 dB for antennas operating across both the N77 and N78 bands. Experimental validation on three representative configurations—a two-element MIMO array, an adjacent-band antenna system, and a multi-element MIMO array—confirms the broadband self-decoupling capability and practical applicability of the proposed technique.

## I. Introduction

IN modern wireless communication terminals, Multiple-Input Multiple-Output (MIMO) antenna technology has become essential to meet the demand for higher channel capacity and diversity performance required by high-speed data transmission, while further improving spectral efficiency and mitigating multipath fading[1]-[3]. However, due to the limited internal space of terminal devices, antenna elements are often arranged in close proximity, leading to severe mutual coupling effects that significantly degrade overall system performance[4]. To address mutual coupling issues, numerous techniques have been proposed, particularly focusing on strong coupling suppression for tightly spaced single-layer microstrip antenna elements [5]-[13] and weak coupling suppression for elements spaced approximately half a wavelength apart [14]-[17]. On the other hand, the growing demand for wider operational bandwidths in terminal devices, especially to support 5G New Radio (NR) bands such as N77 (3.3-4.2 GHz) and N78 (3.3-3.8 GHz), poses significant challenges for conventional single-layer microstrip antennas, whose inherent narrow bandwidth limits their applicability in broadband scenarios. Stacked microstrip patch antennas (S-MPA) have been widely employed to address the stringent bandwidth demands of 5G wireless communication terminals, owing to their superior bandwidth performance and stable radiation characteristics [18]-[24].

At present, the decoupling studies of S-MPA arrays are primarily focused on weak mutual coupling scenarios where the element spacing is around half a wavelength or larger. Common approaches typically rely on the introduction of additional decoupling structures. For example, a combination of the shorting C-shaped structures and square loops was employed in [19], while [20] utilized a combination of symmetric strips and dual-slot single-patch structures. In [21], a full mutual coupling path suppression was achieved through a combination of laminated resonators (LR), coupled line resonators (CLR), and defected ground structures (DGS). Although these methods achieve broadband weak coupling suppression, they tend to involve complex structures with limited robustness, and the relatively large footprint of the decoupling components makes them unsuitable for strong coupling suppression under closely spaced conditions. In [22], a combination of spoof surface plasmon polaritons (SSPPs) and DGS was introduced to achieve size reduction and mutual coupling suppression for S-MPA elements. However, the antenna structure becomes more complicated, and the introduction of DGS increases back radiation and causes E-plane pattern deflection. More critically, such effects are undesirable for many terminal device applications, as they may interfere with components mounted

Manuscript received XX XX 2025; revised X XX 2025; accepted XX XX 2025. Date of publication XX XX 2025; date of current version XX XX 2025. This work was supported in part by the National Natural Science Foundation of China (NSFC) under Grant 62471125. The author received support from the Postgraduate Research & Practice Innovation Program of Jiangsu Province, SJCX25_0076. (*Corresponding author*: Zhen-Guo Liu).

Shao-Hua Xing, Zhen-Guo Liu, Chao Zhang, and Yi-Hao Liu are with the State Key Laboratory of Millimeter Waves, School of Information Science and Engineering, Southeast University, Nanjing 210096, China. (e-mail: liuzhenguo@seu.edu.cn).

Color versions of one or more figures in this article are available at xxxxxxxxxx.

Digital Object Identifier xxxxx

on the opposite side of the system board [9]. In [23], two shorting vias were used to split the driven patch into two regions to achieve mutual coupling suppression; however, this approach adversely affects the boresight gain and impedance bandwidth and does not provide decoupling between nonadjacent elements. In [24], decoupling was realized by optimizing the dielectric constant of the substrate to modify the effective spatial permittivity, which imposes practical limitations on fabrication.

It is worth noting that the aforementioned decoupling methods [19]-[24] achieve the mutual coupling nulls by inducing opposite surface current distributions between the parasitic and driven patches within the frequency band of interest, thereby achieving decoupling. Nevertheless, these methods inherently exhibit high design complexity, sensitivity to fabrication tolerances, low robustness, and a lack of targeted design guidelines. Moreover, the opposite surface currents between the parasitic and driven patches often degrade the boresight gain and total efficiency, and may even introduce unintended in-band radiation nulls [25]. Most importantly, these techniques are only effective for weak mutual coupling suppression when the inter-element spacing is greater than or equal to half a wavelength. In particular, the suppression of strong mutual coupling in densely packed S-MPA arrays remains largely unexplored. When the element spacing is much smaller than half a wavelength—as commonly found in smart terminals and industrial IoT devices—the coupling strength becomes extremely high, rendering traditional S-MPA decoupling techniques based on weak-coupling assumptions ineffective. Furthermore, studies on inter-frequency decoupling within S-MPA arrays (e.g., one element operating in the 5G NR N77 band and another in the N78 band) are extremely limited. Adjacent frequency band coupling involves more complex electromagnetic interference mechanisms and is significantly more challenging to suppress. Consequently, current methods exhibit significant limitations in addressing both strong intra-band and inter-band coupling in compact S-MPA array designs, highlighting the urgent need for novel, efficient, and robust decoupling techniques.

A novel self-decoupling mechanism enabled by a locally strong reverse auxiliary current path is proposed for MIMO S-MPA pair applications. By introducing an embedded ultra-narrow metallic coupling line (EUNMCL) between adjacent parasitic patches, strong mutual coupling can be effectively suppressed even when the driven and parasitic radiating patches exhibit surface current distributions in the same direction. This approach tackles the challenge of mutual coupling suppression between tightly spaced, sub-half-wavelength S-MPA elements, achieving significant isolation enhancement without degrading radiation performance. To the best of the authors' knowledge, this study is the first to propose an effective self-decoupling strategy for S-MPA arrays with co-directional surface current distributions. Furthermore, the proposed method also enables, for the first time, effective self-decoupling between hetero-frequency S-MPA pairs operating in adjacent frequency bands. Compared with conventional decoupling techniques, the proposed self-decoupled S-MPA technology offers the

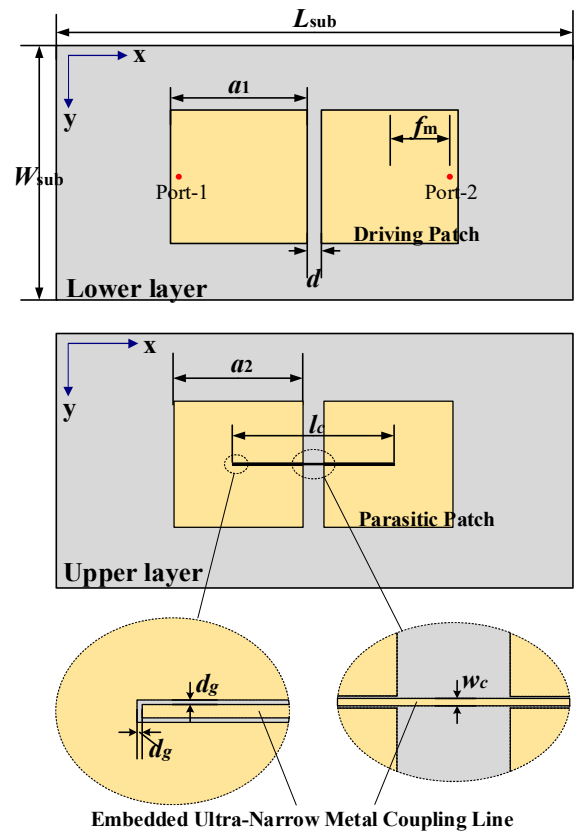

Fig. 1. Geometry of the proposed self-decoupling S-MPA Pair. $a_1$ = 28.7 mm, $f_m$ = 12.6 mm, $d$ = 3 mm, $L_{sub}$ = 150 mm, $W_{sub}$ = 80 mm, $a_2$ = 27.3 mm, $l_c$ = 34.0 mm, $w_c$ = 0.3 mm, $d_g$ = 0.1 mm.

following advantages and distinctive features:

1) Broadband Self-Decoupling: Intrinsic high isolation is achieved across the entire N78 band without additional decoupling circuits or complex fabrication processes;

2) Adjacent-band Decoupling Support: Effective suppression of mutual coupling is realized not only within N78 but also between N77 and N78 bands;

3) Flexible Feeding Locations: The feeding positions on the patches are relatively flexible without stringent placement constraints;

4) High Robustness and Low Cost: The decoupling relies solely on patch geometry, tolerating fabrication inaccuracies and enabling cost-effective mass production;

5) Simple and Intuitive Design: Straightforward design guidelines facilitate easy implementation in practical engineering;

6) Good Scalability: The technique can be extended to multi-element arrays, improving isolation between both adjacent and non-adjacent elements;

7) Broad Element Spacing Adaptability: The method is applicable to both closely spaced and conventionally spaced arrays.

This paper begins with the design of a tightly spaced self-decoupled S-MPA pair. The antenna pair's performance is first characterized. Based on this topology, the self-decoupling mechanism of the proposed method is investigated in Section II-C. Subsequently, key design parameters are studied in

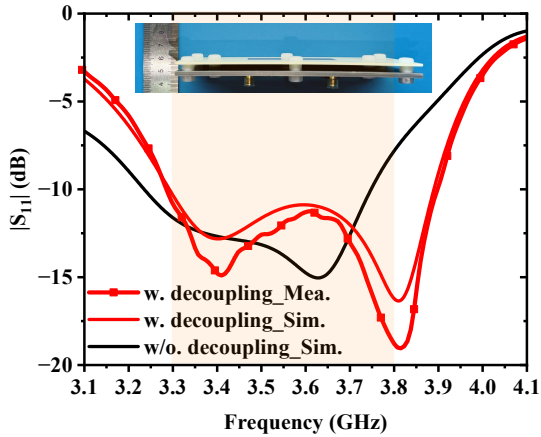

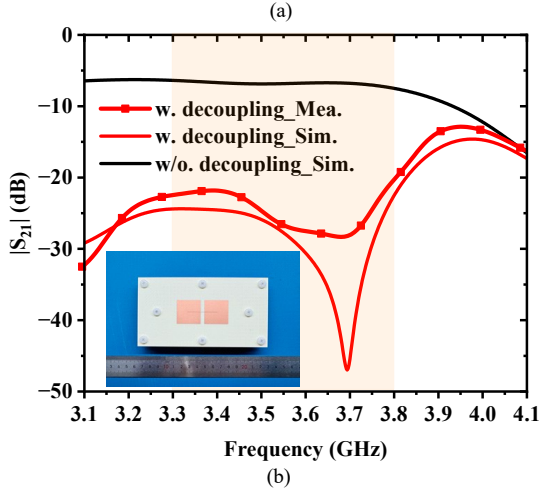

Fig. 2. Simulated and measured *S*-parameters for the proposed self-decoupling S-MPA Pair. (a) Reflection coefficients. (b) Isolations.

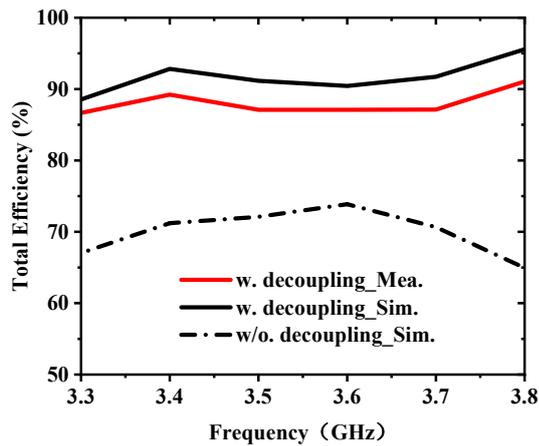

Fig. 3. Simulated and measured total efficiency of the S-MPA Pairs.

Section II-D according to the identified mechanism. A detailed design guideline is then provided in Section II-E. In Section III, potential applications in multi-element array extension and inter-band self-decoupled antenna pairs are discussed through representative examples. A comparison between the proposed technique and state-of-the-art decoupling methods is presented

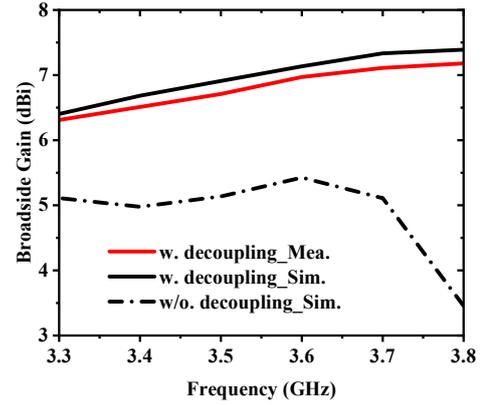

Fig. 4. Simulated and measured broadside gain of the S-MPA Pairs.

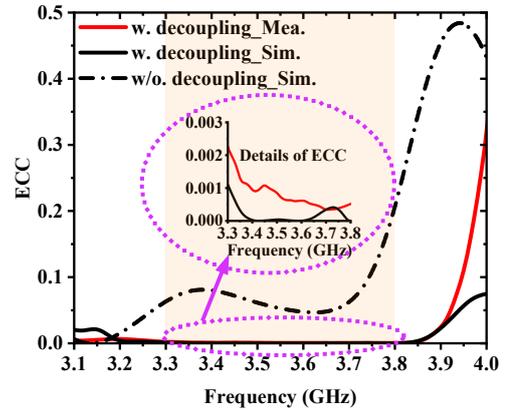

Fig. 5. Simulated and measured ECCs for the S-MPA Pairs.

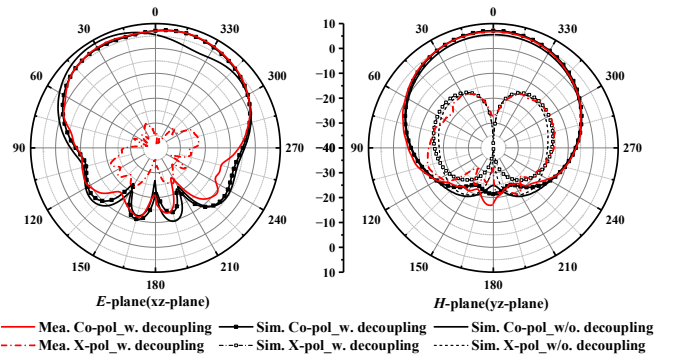

Fig. 6. Simulated and measured radiation patterns of the S-MPA Pairs.

in Section III-C. Finally, concluding remarks are given in Section IV.

## II. Two-Element MIMO Array

### A. Basic Physical Structure

Fig. 1 illustrates the physical configuration of the proposed self-decoupled stacked antenna Pair. In this design, the driven patches are printed on the upper surface of the lower layer (driven layer) substrate, which is implemented using a Rogers 5880 substrate with a dielectric constant of 2.2, a loss tangent of 0.001, and a thickness of 3 mm. The parasitic patches are printed on the upper surface of the upper layer (parasitic layer) substrate, which employs a Rogers RO4350B substrate with a dielectric constant of 3.66, a loss tangent of 0.0037, and a thickness of 1.524 mm. Between the substrates, a 4.2 mm air gap is introduced to enhance the bandwidth and total efficiency of the antennas [8],[9],[21]. The edge-to-edge separation between the two adjacent driven patches is 3.0 mm ($0.035\lambda_c$), where $\lambda_c$ denotes the free-space wavelength at the center frequency of the antenna. An embedded ultra-narrow metallic line (EUNML) with a width of 0.3 mm is inserted between adjacent parasitic patches, in order to avoid the direct connection between the metallic line and the parasitic patch, a gap with a width of 0.1 mm is introduced in the section where the line is embedded within the patch. The 50-Ω coaxial probes are used to feed the driven patches. The proposed antenna Pair can be processed with multilayer substrates through Printed Circuit Board (PCB) process. Full-wave simulations are conducted using electromagnetic field simulation software, CST Studio Suite 2025. The specific design guidelines governing the patch dimensions will be detailed in Section II-E.

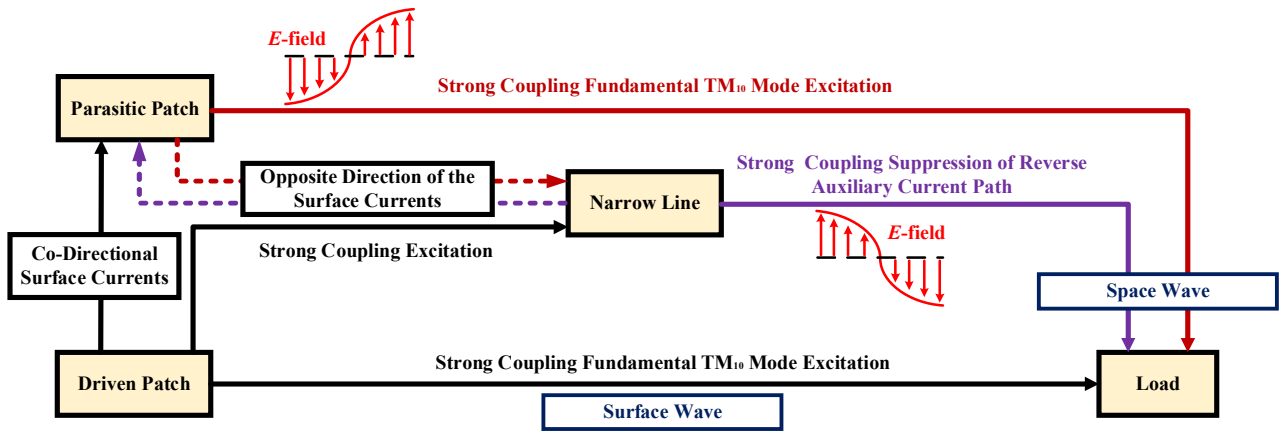

Fig. 7. Schematic Illustration of the Self-Decoupling Mechanism in the Proposed S-MPA Pair.

*B. Antenna Performance*

To validate the effectiveness of the proposed broadband self-decoupling technique for the S-MPA Pair, a prototype antenna, as illustrated in Fig. 1, was fabricated and measured. Fig. 2 compares the *S*-parameters of the proposed self-decoupled S-MPA Pair with those of the conventional S-MPA Pair, along with a photograph of the fabricated prototype. Owing to the symmetrical structure of the antenna Pair, only the case where Port-1 is excited and Port-2 is terminated with a 50 Ω load is presented for simplicity. The measured results demonstrate that the proposed antenna achieves a −10 dB impedance bandwidth of 16.71% (3.29 GHz - 3.89 GHz), fully encompassing the 5G New Radio (NR) N78 band. Within this band, the isolation is enhanced from 7.1 dB to over 20.28 dB (22.1 dB simulated), reaching a peak value of 28.3 dB(46.98 dB simulated), corresponding to an overall improvement of 13.18 dB(15.0 dB simulated). Moreover, the simulated and measured results show strong agreement.

In conventional closely spaced antenna pairs, strong mutual coupling induces unintended excitation of the adjacent antenna when one antenna is excited[5],[6],[8]. This leads to significant energy loss and severely degrades the independent data stream transmission between antenna elements [9],[13]. Consequently, the total efficiency deteriorates, and the radiation pattern is adversely affected.

By implementing the proposed self-decoupling technique, the total efficiency within the N78 band improves from 73.8% to over 86.7% (88.5% simulated), achieving an overall enhancement of 12.9%, as shown in Fig. 3. Additionally, the boresight gain increases from a maximum of 5.4 dBi to over 7.2 dBi (7.4 dBi simulated), as shown in Fig. 4. Besides, to assess the diversity performance of the proposed self-decoupled S-MPA Pair, the envelope correlation coefficient (ECC) curves are plotted in Fig. 5. Both simulated and measured ECC values remain below 0.0022 throughout the entire N78 band, confirming the antenna's excellent diversity performance. In addition, the reference antenna (without decoupling) exhibits significant beam deflection in the *E*-plane due to strong mutual coupling, as observed in Fig. 6. After decoupling, the 3-dB beamwidth increases from 55.7° to 96°(96.2° simulated), indicating a substantial improvement in beam deflection. In the *H*-plane, the boresight gain increases from 5.52 dBi to 6.78 dBi(7.13 dBi simulated) after decoupling, and the cross-polarization performance is also improved. As no modifications are made to the ground plane, the antenna exhibits very low back radiation[5],[9],[14]. Furthermore, this decoupling method induces only minor changes to the original patch structure, ensuring that all antenna elements maintain excellent fundamental $TM_{10}$ mode radiation performance.

*C. Decoupling Mechanism*

To further elucidate the self-decoupling mechanism of the proposed method, Fig. 7 presents the working principle of a tightly spaced self-decoupled S-MPA pair. The driven patch strongly couples to both the corresponding parasitic radiating patch and the embedded narrow line, thereby simultaneously exciting the fundamental $TM_{10}$ mode in both structures. Due to the shorter coupling path to the parasitic patch compared to the narrow line, the induced currents exhibit a phase difference. As a result, the surface currents on the parasitic patch and the embedded narrow line can be consistently out of phase.

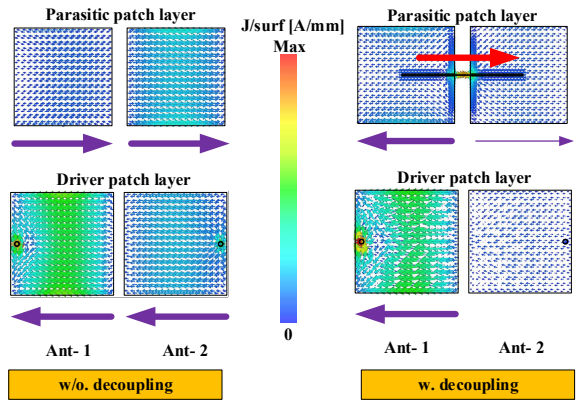

Fig. 8. Comparison of surface current distribution of different layers.

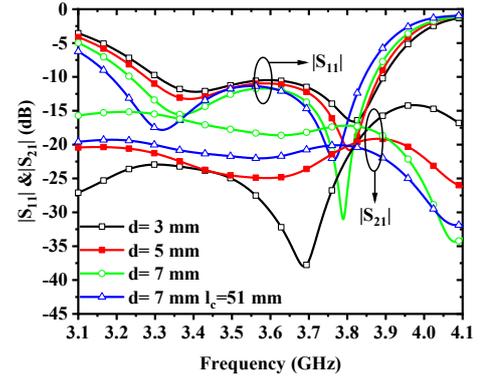

Fig. 10. Simulated *S*-parameters of the Proposed S-MPA Pair under Various Edge-to-Edge Spacing *d*.

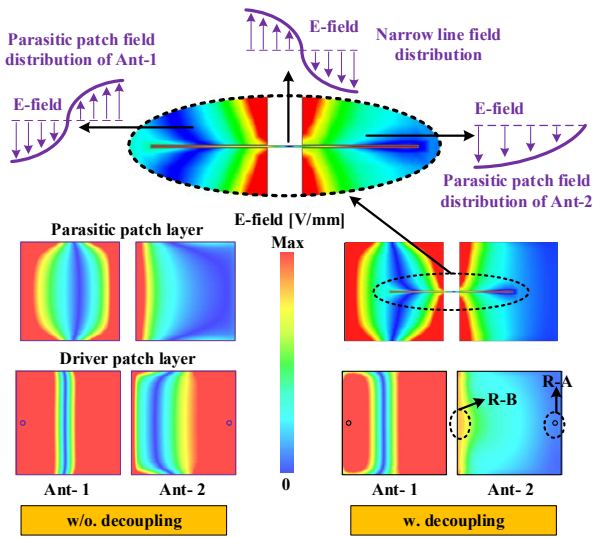

Fig. 9. Simulated *E*-field distribution comparison.

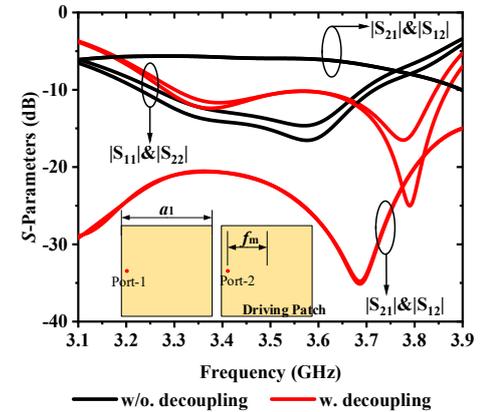

Fig. 11. Comparison of *S*-parameters of conventional *E*-plane-coupled 1 × 2 MIMO S-MPA array. $a_1$ = 29.2 mm, $f_m$ = 13.8 mm, $d$ = 3 mm, $L_{sub}$ = 150 mm, $W_{sub}$ = 80 mm, $a_2$ = 28.1 mm, $l_c$ = 32.1 mm, $w_c$ = 0.3 mm, $d_g$ = 0.1 mm.

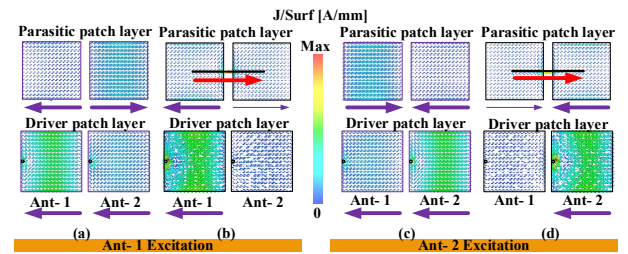

Fig. 12. Comparison of surface current distribution of different layers.

By adjusting the length of the narrow line to an appropriate value, the surface currents on the driven and parasitic patches can be aligned in the same direction, thereby enabling high-gain cooperative radiation between the two radiators. In this state, the driven and parasitic patches directly couple the fundamental $TM_{10}$ mode to the load in-phase via surface and space waves, respectively, while the narrow line contributes the fundamental $TM_{10}$ mode out-of-phase. The superposition of these three coupling paths results in effective mode vector cancellation at the load. Since the narrow line is sufficiently thin and does not serve as an effective radiator [8],[9], it functions solely as a passive coupler that adjusts the surface current distribution on the parasitic patch and contributes to the final mode cancellation at the load. This design not only suppresses strong mutual coupling in the S-MPA system, but also fundamentally avoids the boresight gain degradation and total efficiency loss caused by the reversed current distribution on active radiators in conventional S-MPA array decoupling techniques [19], [20], [21], [24].

Fig. 8 compares the surface current distributions on different layers of the antenna pairs, with and without the proposed self-decoupling technique. As observed, in the case without decoupling, the parasitic and driven patches of Ant-1 exhibit anti-parallel current distributions. Since the inherent mutual coupling null typically lies outside the frequency band of interest [21], [24], this configuration results in relatively low isolation, as also illustrated in Fig. 2(b). Under these conditions, both the driven and parasitic layers of Ant-2 show strong induced currents. After introducing the proposed decoupling method, the driven and parasitic patches of Ant-1 display co-directional surface currents, while the narrow line exhibits an oppositely directed current. Therefore, the induced current on the load is significantly reduced after decoupling, as compared to the case without decoupling. This outcome aligns well with the self-decoupling mechanism described in Fig. 7.

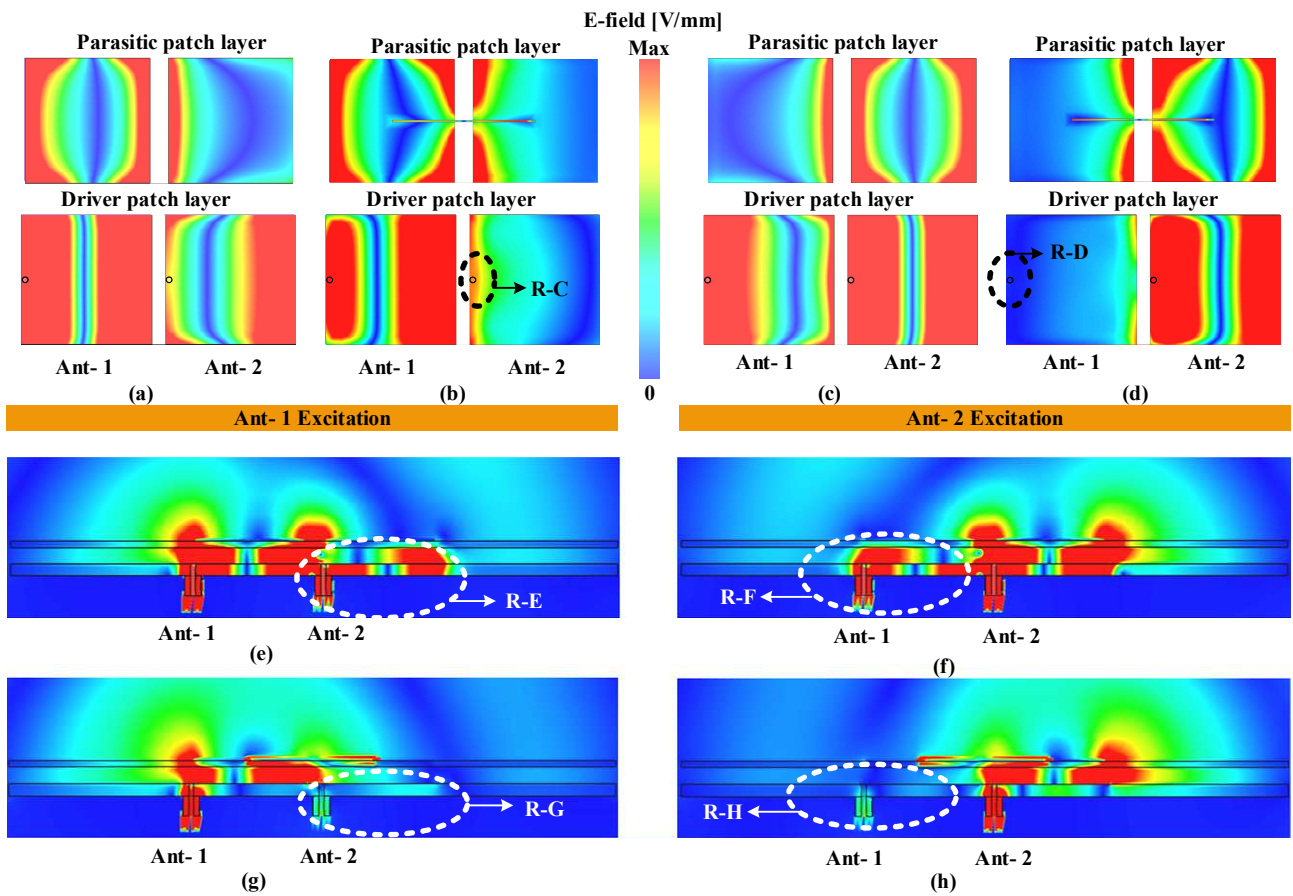

Fig. 13. Simulated *E*-field distribution in plot maximum field amplitude comparison. (a), (c) coupled antenna Pair under Ant-1 and Ant-2 excitation; (b), (d) Proposed self-decoupled antenna Pair under Ant-1 and Ant-2 excitation; (e), (g) Cross-sectional field distribution of coupled antenna Pair under Ant-1 and Ant-2 excitation; (f), (h) Cross-sectional field distribution of Proposed self-decoupled antenna Pair under Ant-1 and Ant-2 excitation.

Fig. 9 illustrates the comparison of *E*-field distributions before and after decoupling. It can be observed that after introducing the narrow line, the fundamental $TM_{10}$ mode on the driven patch of Ant-2, which was originally excited by strong mutual coupling, is completely suppressed. The *E*-field intensity is significantly reduced, and at this time, the feed probe is located in a weak field R-A. In contrast, the driven and parasitic patches of Ant-1 operate in the fundamental $TM_{10}$ mode both before and after decoupling. Although the *E*-field distribution of the narrow line and the parasitic patch of Ant-1 exhibits the same radiation mode, a pronounced difference in field intensity appears in their overlapping region. This can be attributed to a 180° phase difference between the two, which aligns with the oppositely directed current distributions shown in Fig. 8. Interestingly, despite the effective radiation mode on the driven patch of Ant-2 being canceled, a localized strong field is still observed near the radiating edge adjacent to Ant-1. This is not unexpected, as the proposed self-decoupling mechanism is not based on wave blocking [10] and thus does not prevent strong coupling between adjacent patches. On the contrary, this phenomenon confirms the occurrence of coupling cancellation in that region. Furthermore, after decoupling, the electric field of Ant-2 is concentrated above the patch and directed vertically downward, diminishing rapidly with increasing distance, and no resonant field distribution is formed.

Fig. 10 presents the *S*-parameter results for various edge-to-edge spacings *d*. As the spacing increases from 3 mm to 7 mm, the antenna's resonance frequency shifts slightly toward the lower end of the band, and the mutual coupling null correspondingly decreases from 3.7 GHz to 3.55 GHz. This frequency shift is attributed to the elongation of the equivalent current path formed by the narrow line and the parasitic patch. Interestingly, mutual coupling between the antenna elements becomes more pronounced with increasing *d*, which contrasts with the conventional expectation that mutual coupling decreases as inter-element spacing increases. However, this observation is reasonable in the context of the proposed self-decoupling mechanism, which relies on the superposition of coupling fields originating from the parasitic radiating patch, the driven patch, and the narrow line. As the spacing *d* increases, the overall coupling field distribution changes accordingly. As a result, the deviation between the narrow line length and its optimal value increases, leading to a gradual degradation in decoupling performance.

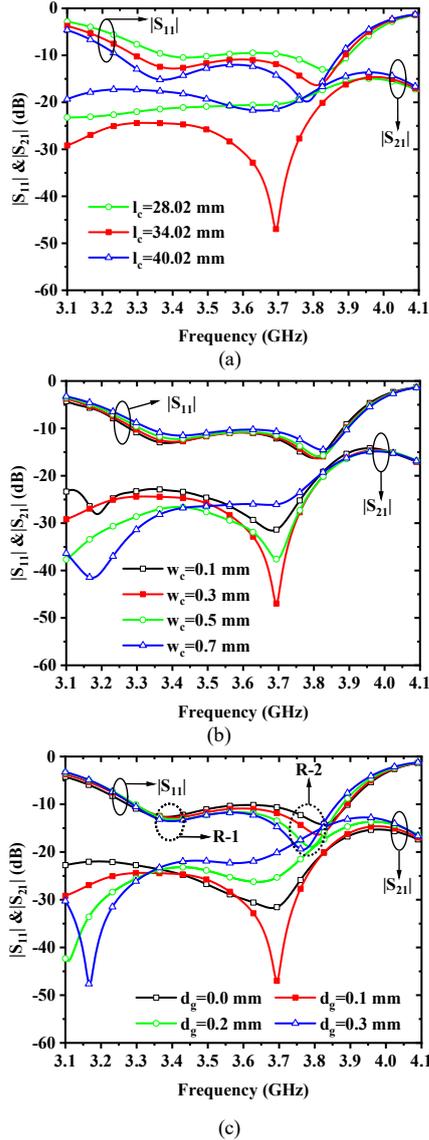

Fig. 14. Simulated *S*-parameters with different (a) length ($l_c$), (b) width ($w_c$), and (c) gap width ($d_g$).

When *d* = 7 mm, optimizing the line length $l_c$ to 51 mm restores the isolation to over 20 dB within the frequency band of interest. Nevertheless, the decoupling depth is still reduced compared to the case of *d* = 3 mm. This is because, as the spacing increases, the electromagnetic interaction between patch elements transitions from strong to weak mutual coupling. In such cases, the suppression of weak mutual coupling in low-profile S-MPA arrays, where the center-to-center distance approaches or exceeds half a wavelength, cannot be effectively achieved by only decoupling the space wave[21]. The residual effects of conduction currents in the ground plane and surface wave in the driven layer remain significant, thereby limiting the decoupling depth[22]. Therefore, the proposed method offers superior performance in suppressing strong mutual coupling for tightly spaced S-MPA pairs in space-constrained environments. For arrays with relatively large inter-element spacing, where there is greater spatial freedom, various alternative techniques can be employed to suppress all coupling pathways effectively[19], [22], [23].

To examine the potential influence of the localized strong *E*-field in the R-B region, as depicted in Fig. 9, on the mutual coupling between adjacent antenna elements, Port-2 was relocated to the R-B position to construct a conventional two-element array aligned along the *E*-plane. The corresponding *S*-parameter responses are shown in Fig. 11. It can be observed that, after decoupling, the isolation within the 5G NR N78 band is significantly improved, increasing from 5.81 dB to 20.61 dB, corresponding to an enhancement of 14.8 dB. Specifically, at 3.68 GHz, the mutual coupling is reduced from -6.67 dB to -34.78 dB, yielding a maximum reduction of 28.11 dB. Furthermore, when Port-1 (Port-2) is individually excited while the other port is terminated with a 50-Ω load, the driven and parasitic patches of Ant-1 (Ant-2) exhibit co-directional surface current distributions. These currents are observed to flow in the opposite direction to that on the narrow line, as illustrated in Fig. 12. This current behavior is in agreement with the self-decoupling mechanism previously outlined in Fig. 7 and Fig.8.

To further elucidate the proposed energy path redirection technique, Fig. 13 presents the simulated electric field distributions, with the maximum field amplitudes plotted for comparison. As anticipated, when Ant-1 and Ant-2 are excited individually, regions R-C and R-D are located within localized strong and weak field regions, respectively. In both scenarios, the load region supports a non-resonant mode, as previously described in Fig. 9.

Notably, as shown in Fig. 13(g) and Fig. 13(h), the electric field intensities at the load-side feed probes (regions R-G and R-H) remain comparably weak in both cases. This observation indicates that, even in the presence of a localized strong electric field, no effective power transmission is established at the feed probe of the load, and a standing wave condition cannot be sustained. As a result, energy cannot be efficiently absorbed. These findings suggest that the original coupling path and associated energy transmission conditions have been fundamentally reconfigured, thereby maintaining a low level of mutual coupling within the antenna system. Consequently, the proposed self-decoupling mechanism is fundamentally distinct from conventional approaches based on electric field suppression—such as electromagnetic bandgap (EBG) structures [26], DGS[22], [27], [28], and asymmetric isolation walls [10]—as well as from methods that place the feed probe in a weak field region [13], [17]. Instead, the present method introduces a new concept based on the redirection of energy paths. As a result, the effectiveness of the proposed self-decoupling technique is not constrained by the specific position of the feed probe.

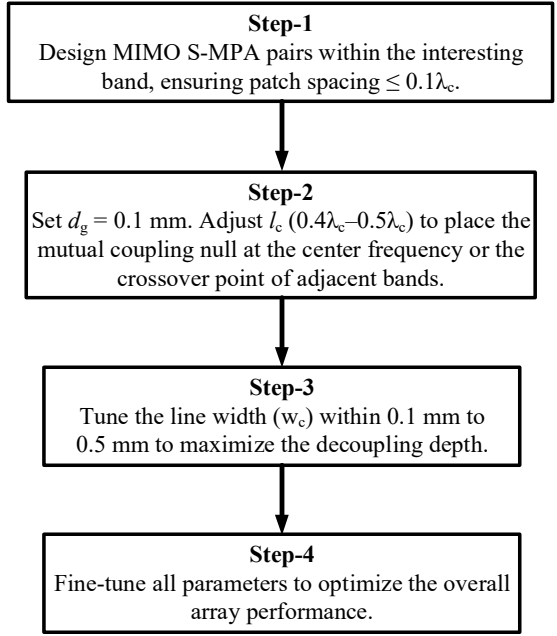

Fig. 15. Design flowchart of the proposed self-decoupling method based on a local strong reverse auxiliary current.

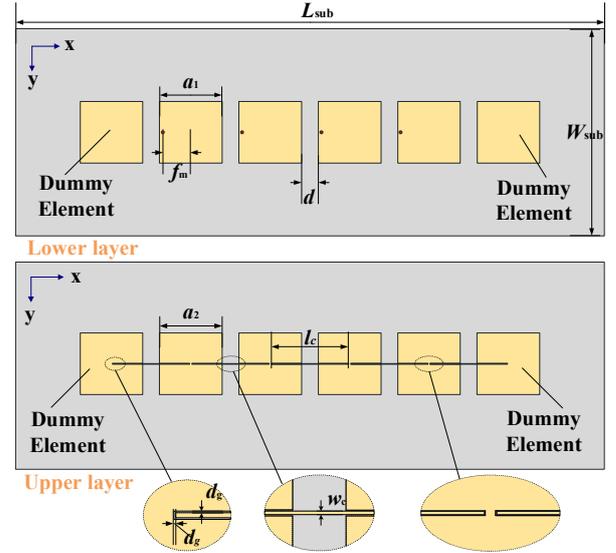

Fig. 16. Dimensions of the 1 × 4 MIMO self-decoupling S-MPA array(unit: mm): $a_1$ = 29.8 mm, $f_m$ = 14.2 mm, $d$ = 8 mm, $L_{sub}$ = 280 mm, $W_{sub}$ = 100 mm, $a_2$ = 29.8 mm, $l_c$ = 36.6 mm, $w_c$ = 0.3 mm, $d_g$ = 0.1 mm.

*D. Parametric Study*

To comprehensively grasp the design concept underlying the proposed self-decoupling technique for the S-MPA Pair, this section investigates several key parameters. During the analysis of the specific parameter, all other parameters are kept constant. Fig. 14(a) illustrates the effect of the narrow line length $l_c$ on the *S*-parameters. As $l_c$ increases, the resonance frequency exhibits a slight shift toward lower frequencies due to the extension of the surface current path on the patch. When $l_c$ increases from 28.0 mm to 34.0 mm, the isolation performance improves significantly, resulting in broadband high-isolation characteristics. However, further increasing $l_c$ to 40.0 mm leads to a degradation in decoupling performance. Therefore, the optimal line length is determined to be 34.0 mm. Fig. 14(b) shows the impact of the narrow line width $w_c$ on the *S*-parameters. The value of $w_c$ has negligible influence on $|S_{11}|$. As $w_c$ increases from 0.1 mm to 0.3 mm, the decoupling depth improves. However, when $w_c$ is further increased to 0.7 mm, a decline in isolation performance is observed. Hence, the optimal width is selected as 0.3 mm. Fig. 14(c) presents the influence of the gap width $d_g$. The gap width $d_g$ has no observable impact on the resonance point R-1 of the driven patch. However, an increase in $d_g$ elongates the effective current path on the parasitic patch, shifting the resonance point R-2 gradually to lower frequencies. In the absence of a gap, a preliminary mutual coupling null appears at approximately 3.7 GHz. Increasing $d_g$ to 0.1 mm results in effective separation of the out-of-phase regions, thereby enhancing isolation. Further increasing the $d_g$ to 0.3 mm causes R-2 to shift further into the lower frequency range while the decoupling depth decreases. Accordingly, the optimal gap width is identified as 0.1 mm.

It is also worth noting that within a fabrication tolerance of ± 0.1 mm around the optimal gap width, the effective decoupling bandwidth ($|S_{11}|$ < −10 dB & $|S_{21}|$ < −20 dB) still fully covers the 5G NR N78 frequency band. This confirms that the proposed structure offers strong dimensional tolerance and high robustness, making it suitable for cost-effective mass production with relaxed fabrication precision requirements.

*E. Design Guideline*

Based on the above analysis, a straightforward design guideline for implementing self-decoupling in tightly spaced MIMO S-MPA pair by embedding a narrow line is summarized, as illustrated in the flowchart provided in Fig. 15.

### III. Application Prospects

In this section, two potential application prospects for the proposed decoupling method will be introduced with two design examples.

*A. Four-Element MIMO Array*

The proposed self-decoupling technique for the S-MPA pair has been further extended to larger-scale MIMO S-MPA arrays for effective mutual coupling suppression. To validate its scalability, a 1×4 antenna array was designed with elements arranged along the *E*-plane. Following the methodology presented in [5], [14], two symmetrically placed dummy elements are added at both ends of the driven and parasitic patch arrays to ensure impedance balance among all elements. Similar to the two-element self-decoupled S-MPA pair, a narrow line with a width of 0.3 mm was inserted between each pair of adjacent parasitic patches, and a gap of 0.1 mm was introduced at the intersections between the line and the parasitic patches. The structural layout of the four-element array is illustrated in Fig. 16, with detailed dimensions annotated in the caption.

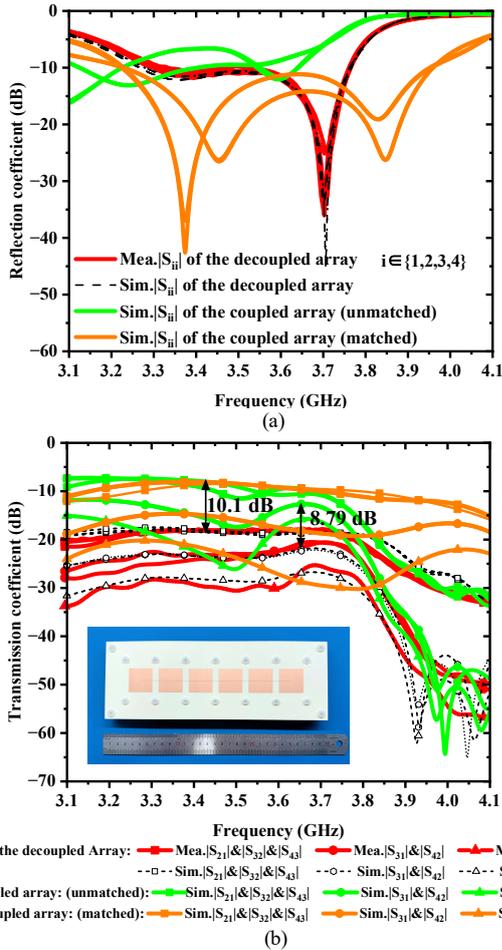

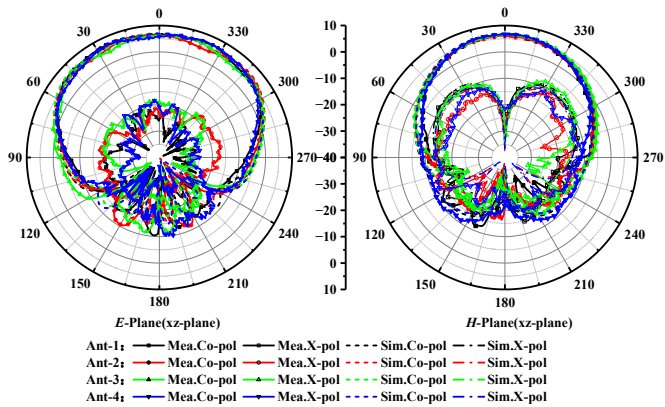

Fig. 17. Simulated and measured $S$-parameters for the 1×4 MIMO self-decoupling S-MPA array. (a) Reflection coefficients. (b) Transmission coefficient.

Fig. 18. Simulated and measured radiation patterns at center frequency.

The array was fabricated and experimentally characterized, as shown in Fig. 17. Measured results indicate that, across the entire 5G NR N78 band, the proposed self-decoupled array achieves isolation levels of at least 18.03 dB (17.46 dB simulated) for adjacent elements ($S_{21}$, $S_{32}$, $S_{43}$), and 20.58 dB (21.81 dB simulated) for non-adjacent elements ($S_{31}$, $S_{42}$). These represent enhancements of 10.01 dB and 8.79 dB, respectively, compared to the coupled antenna array, which exhibits isolation levels of 8.02 dB and 11.79 dB. In contrast, the conventional coupled array experiences significant impedance mismatch across the interesting band due to strong mutual coupling. Although impedance matching can be partially improved by adjusting the dimensions of the driven and parasitic patches, the isolation after matching remains limited to 8.05 dB and 14.68 dB for adjacent and non-adjacent elements, respectively. Moreover, although the spacing between Port-1 and Port-4 is relatively large, which reduces the effectiveness of the self-decoupling mechanism, the proposed array still achieves an isolation improvement of 5.89 dB compared to the coupled array, whose isolation is 19.44 dB, resulting in a measured isolation of at least 25.33 dB (26.81 dB simulated). A photograph of the fabricated four-element array is shown in Fig. 17(b).

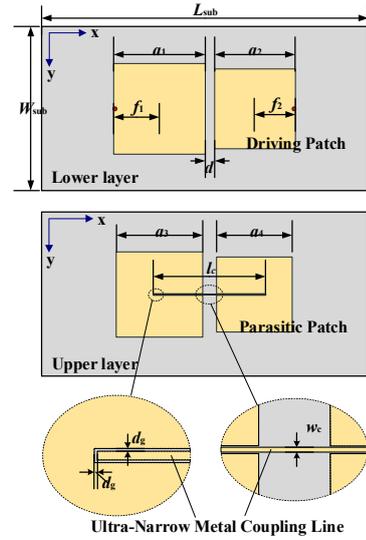

Fig. 19. Geometry of the proposed self-decoupling S-MPA Pair with different operating bands. $a_1$ = 29.1 mm, $a_2$ = 25.7 mm, $a_3$ = 27.7 mm, $a_4$ = 24.3 mm, $f_1$ = 14.1 mm, $f_2$ = 12.4 mm, $d$ = 3 mm, $l_c$ = 35.9 mm, $w_c$ = 0.3 mm, $d_g$ = 0.1 mm, $L_{sub}$ = 150 mm, $W_{sub}$ = 80 mm.

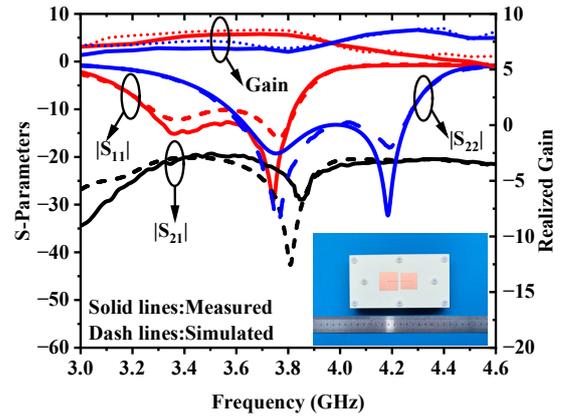

Fig. 20. Simulated and measured $S$-parameters and gain for the proposed self-decoupling S-MPA Pair with the different interesting bands.

Fig. 18 illustrates the measured and simulated radiation patterns of the fabricated four-element self-decoupled S-MPA array. At the center frequency, Ant-1 through Ant-4 exhibit symmetrical radiation patterns in both the $E$-plane and $H$-plane,

TABLE I
COMPARISON OF STATE-OF-THE-ART DECOUPLING TECHNIQUES

| Ref. | Decoupling Method & Central Frequency | Design Complexity | Antenna types | ES* ($\lambda_c$) | Array config. | Non-adjacent elements Decoupling Support | Adjacent-band Operation Support | Broadside Gain Enhancement | IBD* (dB) | IAD* (dB) | OIE* (dB) |
|---|---|---|---|---|---|---|---|---|---|---|---|
| [5] | Mode Cancellation @2.45 GHz | Low | MPA | 0.02 | E- | N.G. | No | Yes | 4 | 14.5 | 10.5 |
| [9] | Modified Patch @5.9 GHz | Low | MPA | 0.002 | H- | N.G. | No | Yes | 6.8 | 18.7 | 11.9 |
| [13] | Self-Decoupling@4.13 GHz | High | Hybrid Modes PA | 0.05 | N.G. | No | Yes | No | 17 | 29 | 12 |
| [17] | CMA @4 GHz | Low | MPA | 0.13 | E- | Yes | No | N.G. | 19 | 30 | 11 |
| [19] | Decoupling structure @3.5 GHz | High | S-MPA | 0.3 | H- | No | No | No | 21 | 29 | 8 |
| [20] | Decoupling structure @3.5 GHz | High | S-MPA | 0.27 | E- | No | No | No | 21 | 31 | 10 |
| [21] | Mutual Coupling Nulls @5.8 GHz | High | S-MPA | 0.5 | E- | Yes | No | No | 24 | 35 | 11 |
| [22] | SSPP&DGS @12 GHz | High | S-MPA | 0.1 | E- | N.G. | No | No | 14 | 21 | 7 |
| [23] | Shorting Pins @3.5 GHz | Low | S-MPA | 0.19 | H- | N.G. | No | N.G. | 23 | 32 | 9 |
| This work | Self-Decoupling @3.5 GHz | Low | S-MPA | 0.036 | E- | Yes | Yes | Yes | 7.1 | 20.3 | 13.2 |

ES*: Edge-to-edge spacing; IBD*: Isolation before decoupling; IAD*: Isolation after decoupling; OIE*: Overall isolation enhancement; N.G: Not given.

along with stable broad-beam characteristics. Specifically, the measured 3-dB beamwidth in the E-plane reaches 114.5° (116.9° simulated). Unlike decoupling techniques that rely on DGS [27], [28], the proposed design maintains an unbroken ground plane, resulting in a lower back-lobe level, enhanced front-to-back ratio, and reduced cross-polarization. The front-to-back ratio reaches 19.71 dB (22.12 simulated) across the frequency band of interest. The measured cross-polarization levels in the E-plane and H-plane are 25.7 dB and 17.6 dB below their corresponding co-polarization levels, respectively.

*B. Adjacent-Band Decoupling*

The proposed self-decoupling technique is also applicable to tightly spaced S-MPA pairs operating in adjacent frequency bands. In practical terminal antenna systems, the 5G NR bands N77 and N78 are spectrally adjacent or even partially overlapping. Due to the limited selectivity of filters, it is impossible to guarantee the complete suppression of adjacent-band interference. Conversely, the employment of high-selectivity filters substantially increases system complexity. Furthermore, the compact size of terminal devices demands sufficient isolation between elements to avoid the transmitted signal in one channel from being received by an adjacent receive channel [8]. This imposes the requirement of a broad decoupling bandwidth to effectively mitigate inter-channel crosstalk. Existing solutions, such as coupling resonators [21], decoupling networks [29], and decoupling surface [30], often encounter layout constraints and added design complexity in closely spaced configurations. By contrast, the proposed technique effectively suppresses mutual coupling between hetero-band antenna elements, while offering advantages in structural simplicity, robustness, and design practicality.

Fig. 19 illustrates the structural configuration of the hetero-frequency self-decoupled S-MPA pair, in which Ant-1 operates in the N78 band and Ant-2 in the N77 band, with an edge-to-edge spacing of 3 mm. The decoupling mechanism follows the same principle as that of the aforementioned same-frequency S-MPA pair, with the only modification being the length of the narrow line, which is adjusted such that the mutual coupling null is shifted to the intersection frequency of the two bands at 3.8 GHz. As shown in Fig. 20, both measured and simulated results confirm that the isolation between Ant-1 and Ant-2 remains above 20 dB across the entire N77 and N78 bands. The measured −10 dB impedance bandwidth of Ant-1 is 3.26-3.82 GHz (simulated: 3.27-3.83 GHz), while that of Ant-2 is 3.56-4.28 GHz (simulated: 3.57-4.28 GHz). A photograph of the fabricated antenna is provided in Fig. 20. Although the impedance bandwidth of Ant-2 (18.95%) only covers the exclusive region (3.8-4.2 GHz) of the N77 bands, the results nonetheless verify the effectiveness of the proposed self-decoupling technique for tightly spaced S-MPA pairs operating in adjacent frequency bands. Furthermore, the realized gains of Ant-1 and Ant-2 within their respective operating bands exceed 7.67 dBi (8.19 dBi) and 6.59 dBi (6.89 dBi), respectively.

*C. Comparison and Discussion*

To highlight the advantages of the proposed decoupling mechanism, Table I compares the proposed solution with current state-of-the-art patch antenna decoupling techniques.

First, the proposed method utilizes a local strong reverse auxiliary current to achieve wideband self-decoupling for the S-MPA pair, where the surface currents on both the driven and parasitic radiation patches are co-directional. This approach effectively mitigates strong mutual coupling while addressing the issue of boresight gain, which is a common limitation in traditional decoupling techniques that rely on opposite current distributions between radiators [19]-[23].

Second, the proposed self-decoupling scheme provides a simple and efficient solution for wideband decoupling in the S-MPA array without the need for additional decoupling structures [19]-[22] or complex fabrication processes and material requirements [24]. It meets high isolation demands across the entire N78 frequency band. The design guidelines provided in Section II-E streamline the decoupling implementation, thus avoiding the extended design cycles often associated with the integration of complex structural combinations in high-isolation antenna systems [13],[19]-[22]. While the S-MPA in Ref. [23] is structurally simple, it is limited to weak mutual coupling suppression, does not support decoupling for adjacent frequency bands, and offers inferior wideband decoupling performance compared to the proposed method.

Third, the proposed S-MPA array decoupling technology is suitable for both co-frequency and adjacent-frequency band decoupling, significantly expanding its potential application range. In contrast, most existing S-MPA methods are restricted to co-frequency decoupling [19]-[24]. While electromagnetic coupling cancellation (EMCC) [8] and the hybrid-mode patch antennas in Ref.[13] effectively address strong mutual coupling in closely spaced patch antennas operating in both the same and adjacent frequency bands, these methods are limited to single-layer patch antennas and are not fully applicable to stacked patch antennas, which require wider decoupling bandwidths and are more susceptible to interference from multiple sources [20],[21].

Fourth, the proposed decoupling technique imposes no strict constraints on the feed probe location and patch arrangement. In contrast to the method in Ref. [8], where radiating patches must be offset along the resonant direction and the feed probes must be positioned accordingly, and the restrictions on probe positions in [11], [13], and [17] due to specific mode excitation and field distributions, the proposed approach offers greater flexibility in feed probe positioning and patch layout. This flexibility enables broader application possibilities.

Fifth, the proposed decoupling structure is highly robust and cost-effective. It achieves wideband decoupling (within the N78 band) with a ±0.1mm variation in the optimal gap width. Compared to single-layer decoupling techniques [5]-[18], it offers a wider decoupling bandwidth and more effective suppression of multiple coupling paths. Unlike the decoupling structures in [19] and [20], the resonator units in [21], and the SSPP structures in [22], which impose strict size constraints and necessitate the use of vias or ground plane slots, the proposed solution is more robust and economically viable.

Sixth, the proposed S-MPA decoupling scheme is scalable to multi-element arrays, enhancing isolation between both adjacent and non-adjacent elements. In contrast, the methods in [19],[20],[22],[23] are limited to decoupling adjacent elements. While Ref. [21] can achieve decoupling of non-adjacent elements, its decoupling structures are overly complex, which reduces design efficiency.

Finally, the proposed decoupling technology supports strong mutual coupling suppression at a spacing of 0.036λ, offering a broader range of applications compared to most S-MPA decoupling methods [19]-[21],[23]. While the method in Ref. [22] can suppress strong mutual coupling at a spacing of 0.1λ, it increases back radiation due to the ground plane slots, which may interfere with devices placed on the opposite side of the system board in many practical applications [9]. In contrast, the proposed method preserves the integrity of the ground plane.

## IV. CONCLUSION

A simple and cost-effective self-decoupling technique for tightly spaced S-MPA arrays has been proposed and validated. By embedding narrow lines between adjacent parasitic patches, strong mutual coupling is effectively suppressed under co-directional surface currents, achieving 20.28 dB isolation in the entire N78 band without degrading radiation performance. The method also enables inter-band decoupling between N77 and N78, with coupling levels below -20 dB. Featuring low cost, high robustness, and good scalability, the proposed technique offers a practical and efficient solution for compact MIMO and in-band full-duplex (IBFD) antenna-in-package (AiP) systems.